\documentclass[12pt]{article}

\def\bx{{\bf x}}

\def\cA{{\cal A}}\def\cB{{\cal B}}
\def\cF{{\cal F}}\def\cH{{\cal H}}
\def\cI{{\cal I}}
\def\cM{{\cal M}}\def\cO{{\cal O}}

\def\a{\alpha}

\def\s{\sigma}\def\t{\tau}

\def\2#1{{}^{(2)}\!#1}\def\3#1{{}^{(3)}\!#1}
\def\4#1{{}^{(4)}\!#1}\def\+#1{{}^+\!#1}
\def\-#1{{}^-\!#1}
\def\*#1{{}^*\!#1}

\newtheorem{prop}{Proposition}
\newtheorem{lem}{Lemma}

\title{\bf Quantum Coordinates of an Event in Local
  Quantum Physics}

\author{R. Giannitrapani \\
  \emph{\normalsize Dipartimento di Fisica dell'Universit\`a di Trento
    (Italy)}}

\begin{document}
\maketitle

\begin{abstract}
  Recently \cite{toller} it has been proposed, using the formalism of
  positive-operator-valued measures, a possible definition of quantum
  coordinates for events in the context of quantum mechanics. In this
  short note we analyze this definition from the point of view of
  local algebras in the framework of local quantum theories.
\end{abstract}


UTF-411.

\section{Introduction}

Local quantum theory is now a well established framework for physical
concepts and theories \cite{haag1,haag2}. A quite interesting problem
is to see whether a given local quantum theory admits a sharp
localization of a spacetime event; this problem is intrinsically
related to the definition of spacetime at small scale and so to the
quantum gravity problem.

In a recent work \cite{toller} Toller has proposed to define the
localizability of an event using quantum coordinates via positive
operator valued measures (POV-measures) \cite{busch1}; such useful
formalism has been used
\cite{holevo,busch2,giannitrapani,atmanspacher} to investigate the
longstanding problem \cite{grot,delgado0,delgado,leon,muga,aharonov}
of time in quantum mechanics and the quantum coordinates of
\cite{toller} are a generalization of these works. In this short note
we want to stress the fact that such quantum observable for the
coordinates of an event cannot be built with quasilocal operators.

\section{POV measures in the local framework}

Let $\cM$ be the Minkowsky spacetime and $\cF(\cM)$ be the
$\s$-algebra of borel subsets of $\cM$; to the physical quantum object
that defines the event is associated an Hilbert space $\cH$ and we
call $\cB(\cH)$ the algebra of bounded linear operators on $\cH$ and
with $\cB(\cH)^+$ the positive ones. A POV measure on $\cM$ with value
in $\cB(\cH)^+$ is a map

$$
\t : \cF(\cM) \longrightarrow \cB(\cH)^+ $$
such that $\t$ is
$\s$-additive in the weak topology, $\t(\emptyset) = 0$ and $\t(\cM)
\leq 1$.  If $\t$ has to fix a quantum event we require that it is
covariant with respect to a unitary representation of the Poincar\'e
group.  We remark the fact that we cannot, in general, require the
normalization $\t(\cM)=1$ for the POV-measure, as discussed in
\cite{toller}.

Let us consider now the formulation of a local quantum theory
\cite{haag2} in terms of local operators algebras; it is defined by a
net of $C^*$-algebras, i.e. by an inclusion preserving assignment to
every open, bounded region $\cO$ of $\cM$ of a unital $C^*$-algebra
$\cA(\cO)$; the closure in the norm topology of their union is the
algebra of quasilocal observables

$$ \cA = \overline{\bigcup_{\cO\subset\cM} \cA(\cO)}.$$

It is quite immediate to see that $\t(\cI)$, with $\cI\in\cF(\cM)$ and
$\cI\neq\emptyset$, is not a local operator, i.e. it does not belong
to one of the local algebras $\cA(\cO)$. To see this we remind the
reader that the vacuum state $\Omega$ is invariant under the
translations on $\cM$ and so

$$
(\Omega, \t(\cI)\Omega) = (\Omega, U(x)\t(\cI) U^{-1}(x)\Omega) =
(\Omega,\t(\cI_x)\Omega) $$
where $U(x)$ is an unitary representation on $\cH$ of the group of
spacetime translations and $\cI_x$ is the region $\cI$ transformed
with the translation by $x$. If for $\cI$ we have

$$(\Omega,\t(\cI)\Omega) = \epsilon $$
with $\epsilon > 0$, it does
exist an integer $n$ such that $n\epsilon > 1$. Now we can chose $n-1$
vectors $\{ x_1, \ldots , x_{n-1} \}$ such that $\cI \cap \cI_{x_i} =
\emptyset$ and $\cI_{x_i} \cap \cI_{x_j} = \emptyset \quad \forall
i,j\quad i\neq j$; for the additivity of $\t$ and the invariance of
the vacuum we have

$$
(\Omega, \t(\cI \cup \cI_{x_1} \cup \ldots \cup
\cI_{x_{n-1}})\Omega) = n\epsilon > 1$$
that is impossible. So
$(\Omega,\t(\cI)\Omega) = 0$ and, since $\t(\cI)$ is bounded and
positive, for the Reeh-Schlieder \cite{reeh} theorem it cannot be a
local operator.

With a little more tricky demonstration we can see that it is
neither a quasilocal operator. To see this we need the following
trivial lemma

\begin{lem}
  Given a POV measure $\t$ on $\cM$ covariant with respect to a
  unitary representation of the translations group, if $\t(\cI)$ is a
  quasi-local operator for some region $\cI$ then it is a quasi-local
  operator for every translated regions $\cI_x$.
\end{lem}

A second necessary lemma is a standard result of local quantum physics
(see \cite{haag2, jost, streater}); let us consider $n$ quasi-local
operators

$$
A_1 \ldots A_n \in \cA$$
and the automorphism of $\cA$ associated
to a translation of $x$ given by $\a_x A \equiv U(x) A U^{-1}(x)$.
Let us divide the set of the indices $\{1\ldots n\}$ in two disjoint
subsets $I$ and $J$ and define $n$ new quasi-local operator $A'_1
\ldots A'_n$ by

$$ A'_i = A_i \quad i\in I$$
$$ A'_i = A_i(x) \quad i\in J$$
where $A_i(x) = \a_x A_i$. If we indicate $x = (t, \bx)$ then we have

\begin{lem}[Cluster property]

  For $|\bx| \rightarrow \infty$ 
  $$
  (\Omega, A'_1 \cdots A'_n \Omega) \rightarrow (\Omega,\prod_{i\in
    I} A'_i \, \Omega) \cdot (\Omega, \prod_{j\in J} A'_j \,
  \Omega).$$
\end{lem}
With these lemmas we can now demonstrate the following 

\begin{prop}
  If $\t$ is a POV measure on $\cM$ covariant with respect to the
  spacetime translations group, then $\t(\cI)\notin \cA$  for all
  $\cI\in\cF(\cM)$ with $\t(\cI) \neq 0$. 
\end{prop}
{\flushleft \em Proof}

Let us suppose that $\t(\cI)$ is a quasi-local operator
and let $\psi$ be a normalized vector such that 

$$
(\psi, \t(\cI) \psi) = p$$
with $ 0 <p \leq 1$; since the subset of
vectors obtained by applying quasi-local operators to $\Omega$ is
dense in $\cH$ one can assume that

$$ \psi = A \Omega$$
where $A$ is a quasi-local operator.
Defining the vector $\psi_x = A(x) \Omega$, where $x = (t,
\bx)$, for the covariance property of $\t$ we have

$$ (\psi_x, \t(\cI_x)\psi_x) = p.$$ 
We can now show that it exists a vector sequence 
$\{\tilde{\psi}_x\}$ such that, for $|\bx| \rightarrow \infty$,

$$ (\tilde{\psi}_x, \t(\cI\cup\cI_x) \tilde{\psi}_x) \rightarrow 2p $$ 
and this is obviously impossible since $\t(\cM) \leq 1$.

In fact let us define the normalized vector

$$ \tilde{\psi}_x = \frac{A(x)A\Omega}{\|A(x)A\Omega\|};$$

using the clustering lemma one can see that 

$$\|A(x)A\Omega\|^2 \rightarrow \|A(x)\Omega\|^2\cdot
\|A\Omega\|^2=1 $$  
and that 

$$ (\tilde{\psi}_x, \t(\cI) \tilde{\psi}_x) \rightarrow 
(\Omega, A^*(x) A(x) \Omega)\cdot(\Omega, A^* \t(\cI) A \Omega)=
(\psi, \t(\cI) \psi) = p$$
for $|\bx|\rightarrow\infty$. Similarly, in the same limit,

$$ (\tilde{\psi}_x, \t(\cI_x) \tilde{\psi}_x) \rightarrow 
(\Omega, A^* A \Omega)\cdot(\Omega, A^*(x) \t(\cI_x) A(x) \Omega)=
(\psi_x, \t(\cI_x) \psi_x) = p.$$

So
$$
(\tilde{\psi}_x, \t(\cI\cup\cI_x) \tilde{\psi}_x) =
(\tilde{\psi}_x, \t(\cI) \tilde{\psi}_x) + (\tilde{\psi}_x, \t(\cI_x)
\tilde{\psi}_x) \rightarrow 2p.$$

If $2p > 1$ the demonstration is over; otherwise we can repeat the
same argument finding normalized states such that the preceding
probability tends toward $4p$ and so on.
\begin{flushright}
  \em Q.E.D.
\end{flushright}

So we see that the existence of a normalized POV measure covariant
with respect to a unitary representation of the group of translations
on $\cM$ is incompatible with the local principle if we require that
$\t(\cI)$ is quasi-local.

\section{Conclusions}

The use of POV-measures as observables is motivated by some deep
considerations for the foundations of quantum mechanics
\cite{giles,ludwig}. In particular it is very interesting the use of
positive bounded operators, not necessarily projectors, as generalized
propositions; it is an open problem to see what kind of observables it
is possible to build with them. In this letter we have shown that a
localizability for events in spacetime cannot be described by one of
these observables if the generalized propositions have to belong to
the algebra of quasilocal operators; further investigations are
necessary in this direction.

\subsection*{Acknowledgments}
I wish to thank Marco Toller whose ideas inspired this work and Valter
Moretti for some useful discussions on the subject.

\end{document}